\documentclass{llncs}

\usepackage {amssymb}
\begin{document}

\title{Logic Functions and Quantum Error Correcting Codes}

\author{Yajie Xu$^1$, Zhi Ma$^1$,~Chunyuan Zhang$^1$,~Xin L\"u$^2$}
\institute{$^1$ Department of Information Research, Information
Engineering University, Zhengzhou, 450002, China
\\e-mail:xyjxwf@yahoo.com.cn\\ $^2$ Informatization Institute,
 State Information Center, Beijing, 100045, China.}

\maketitle

\begin{abstract}
In this paper, based on the relationship between logic functions and
quantum error correcting codes(QECCs), we unify the construction of
QECCs via graphs, projectors and logic functions. A construction of
QECCs over a prime field $F_p$ is given, and one of the results
given by Ref\cite{SQC07} can be viewed as a corollary of one theorem
in this paper. With the help of Boolean functions, we give a clear
proof of the existence of a graphical QECC in mathematical view, and
find that the existence of an $[\kern-0.15em[n,k,d]\kern-0.15em]$
QECC over $F_p$  requires similar conditions with that depicted in
Ref\cite{DRF02}. The result that under the correspondence defined in
Ref\cite{MN}, every $[\kern-0.15em[n,0,d]\kern-0.15em]$ QECC over
$F_2$ corresponding to a simple undirected graph has a Boolean basis
state, which is closely related to the adjacency matrix of the
graph, is given.

After a modification of the definition of operators, we find that
some QECCs constructed via projectors depicted in Ref\cite{VR07} can
have Boolean basis states. A necessary condition for a Boolean
function being used in the construction via projectors is given. We
also give some examples to illustrate our results.

\end{abstract}

\section{Introduction}
Quantum error correcting codes(QECCs)have have received more and
more attention for nearest few decades since the theory of quantum
error correction was put forward\cite{AM96,AP96,P98,AEPN98}. One of
the central tasks in the theory of QECCs is how to construct them,
and the first systematic mathematical construction is given
in\cite{AEPN98} in the binary case and then generalized in
\cite{RT00,AE01}. Many good non-binary QECCs have been constructed
by using classical error-correcting codes over $\mathbb{F}_q$ or
$\mathbb{F}_{q^2}$ ($q$ is a power of an odd prime ) with special
orthogonal properties. Besides these, many quantum codes are
constructed via tools including graphs and Boolean
functions\cite{LED05,SQC07,DRF02,MZ}. In this paper, we unify the
two tools.

In Ref\cite{LED05}, the author constructed
$[\kern-0.15em[n,0,d]\kern-0.15em]$ QECCs using Boolean functions
with $n$ variables and aperiodic propagation criterion(APC) distance
$d$, and gave an algorithm to compute the APC distance of a Boolean
function, orbits of Boolean functions of the same APC distance are
also studied. In this paper, a construction of quantum codes of
dimension more than 0 is given basing on the relationship between
logic functions and quantum codes. In Ref\cite{SQC07}, the authors
constructed quantum codes by giving $K$ basis states based on a
graph state associated with a graph, and each basis state
corresponds to a subset of the vertexes, which can be viewed as a
corollary of this paper with weaker requirements.

In Ref\cite{DRF02}, the authors initiated the construction of QECCs
 via the the construction of matrixes with some properties and proved
 the necessary and sufficient conditions for the graph that
  the resulting code corrects a certain number of errors in physics view. In Ref\cite{MZ},
  the author gave another proof in mathematical view. Based on the
  mathematical proof and the close relationship between logic functions and
  matrixes, properties that a logic function should
  have in order to construct a QECC is systematically studied.and
find that the existence of an $[\kern-0.15em[n,k,d]\kern-0.15em]$
QECC over $F_p$  requires similar conditions with that depicted in
Ref\cite{DRF02}.  Under the correspondence defined in Ref\cite{MN},
we find that every $[\kern-0.15em[n,0,d]\kern-0.15em]$ over $F_2$
corresponding to a simple undirected graph with the adjacency matrix
$\Gamma_{n\times n}$ has the basis state
$|\psi_f\rangle=2^{-\frac{n}{2}}\sum\limits_{x\in
F^{n}_2}(-1)^{f(x)}|x\rangle$, where $f(x)=\frac{1}{2}x\Gamma x^T$.
We also give an example to illustrate our result.

In Ref\cite{VR07}, the author describes a common mathematical
framework for the design of QECCs basing on the correspondence
between Boolean functions and projection operators. We point out
that in some conditions, the basis states of the QECCs under the
framework can have the probability  vector of the form
$2^{-\frac{n}{2}}(-1)^{f(x)}$, where $f(x)$ is a Boolean function.

\section{Preliminaries}
 We often express an error ( also can be viewed as an
operator) operating on $\mathbb{C}^{q^n}$ as $E_{(a,b)}=i^\lambda
\mathcal {X}_a \mathcal {Z}_b$, where $a,b$ are vectors of length
$n$ over $F_q$, where $q$ is a power of a prime $p$. And
\begin{equation}
E_{(a,b)}|x_1x_2 \cdots x_n\rangle=i^\lambda
\zeta^{tr_{q/p}(\sum\limits_{j=1}^{n}b_jx_j)}|x'_1x'_2 \cdots
x'_n\rangle
\end{equation}
where $x'_{j}=x_{j}+a_j$, $\zeta$ is a p-th primitive root of 1.

Especially, an error acting on a n-qubit state
 in $\mathbb{C}^{2^n}$ has simpler forms .

\begin{definition}
Operators $E_{(a,b)}$ associated with binary vectors $(a,b)\in
F_2^{2n}$ are defined by
\begin{equation}
E_{(a,b)}=e_1\otimes \ldots \otimes e_{m}=i^{a\cdot b}\mathcal {X}_a
\mathcal {Z}_b
\end{equation}
where $e_{i}=\left\{ \begin{array}{l}
 I_2,\ \ \ a_i=0,b_{i}=0. \\
 \sigma_{x},\ \ \ a_{i}=1,b_{i}=0.  \\
 \sigma_{z} ,\ \ \ a_{i}=0,b_{i}=1.\\
 \sigma_{y},\ \ \ a_{i}=1,b_{i}=1.
 \end{array} \right.$.

\begin{definition}
The weight of an error $E_{(a,b)}=i^\lambda \mathcal {X}_a \mathcal
{Z}_b$ is defined by the symplectic weight of two vectors $a,b$ of
length $n$, i.e.,
\begin{equation}
W_{s}(a,b)=\sharp \{ i| 1\leq i \leq n,(a_{i},b_{i})\neq (0,0)\}
\end{equation}

\end{definition}

We generalize the definition of APC (aperiodic propagation
criterion)distance of a Boolean function in Ref\cite{LED05} to
$F_p$.
\begin{definition}
The APC (aperiodic propagation criterion)distance of a logic
function $f$ over $F_p$ is defined by the smallest nonzero
$w_{s}(a,b)$, where $a,b\in F^{n}_p$ such that
\begin{equation}
\sum\limits_{x\in F^{n}_p}\zeta^{f(x)+f(x-a)+b\cdot x}\neq 0
\end{equation}
where $\zeta$ is a $p$-th primitive root of 1.
\end{definition}

Let $G=(V,E)$ be a graph with vertex set $V=X\cup Y$ and edge set
$E=V\times V$. Each edge $\overline{uv}\in E(u,v \in V ) $ is
assigned a weight $a_{uv}(=a_{vu})\in F_{p}$ . Therefore, such a
graph $G$ corresponds to a symmetric matrix over $F_{p}$ $ A _{n
\times n}= (a_{uv})_{u,v\in V}$ .

For two subsets $S$ and $T$ of $V$, we denote $A_{S, T}$ as the
submatrix of $A$ with size $|S|\times|T|$

\begin{center}
 $A_{S, T} =  (a_{uv})_{u\in S,v\in T }$
\end{center}

Similarly, a vector in the vector space $F_{p}^{|V|}=F_{p}^{n+k}$ is
denoted by a column vector
\begin{center}
\[d^{V}= \left( {\begin{array}{*{20}c}
   {d^{v_1 } }  \\
   {d^{v_2 } }  \\
    \vdots   \\
   {d^{v_{n + k} } }  \\
\end{array}} \right)=\{d^v|v\in V\}
\]
\end{center}
where $d^{v_{i}} \in F_p$. For a subset $S$ of $V$ , we denote
 $d^{S} = \{d^s|s\in S\}\in F_{p}^{S}$,  and $O^{S}$ a vector
of length $|S|$ with every coordinate equal to 0.

Let $E$ be a subset of $Y$ with $d-1$ elements, $I = Y \setminus E$,
then

\begin{center}
\[
A=\left( {\begin{array}{*{20}c}
   A_{XX} & A_{XE} & A_{XI}  \\
   A_{EX} & A_{EE} & A_{EI}  \\
   A_{IX} & A_{IE} & A_{II}  \\
\end{array}} \right)
\]
\end{center}

\begin{lemma}\cite{DRF02}
Suppose $X,Y$ are two disjoint sets. $|X|=k,|Y|=n\geq k+2d-2,d\geq
2$, $A=(a_{ij})_{i,j\in X\cup Y}$ is a symmetric matrix with
 vanishing diagonal entries.  For
arbitrary $E\subseteq Y,|E|=d-1$, if

\begin{equation}
A_{IX}d^X+A_{IE}d^E=O^I
\end{equation}

with $I=Y \setminus E $ implies that

\begin{equation}
d^X=O^X~ and ~A_{XE}d^E=O^X
\end{equation}
Then there exists an $[\kern-0.15em[n,k,d]\kern-0.15em]$ quantum
code.
\end{lemma}

\begin{definition} \cite{VR07} 
We define the $Zset_{f}$ of a Boolean function $f$ by
\begin{center}
$Zset_{f}=\{a|\sum\limits_{x\in F_2^{n}}f(x)f(x+a)=0\}$
\end{center}
\end{definition}
\end{definition}

\begin{lemma}\cite{VR07}
If the weight of the Boolean function $f$ with $n$ variables is $M$,
and $M\leq 2^{n-1}$, then $Zsetf=\{a|r_f(a)=2^n-4M\}$, where
$r_f(a)$ is the autocorrection function of $f(v)$ at $a$, i.e.,
$r_f(a)=\sum\limits_{x\in F_2^{n}}(-1)^{f(x)+ f(x+ a)}$.
\end{lemma}

\begin{lemma}\cite{VR07}
An $((n,M,2))$-QECC is determined by a Boolean function $f$ with the
following properties

\begin{itemize}
 \item[1)] $f$ is a function of $n$ variables and has weight $M$.
 \item[2)]The $Zset_f$ contains the set
$\{\alpha_1,\alpha_2,\ldots,\alpha_{2n},\alpha_1+\alpha_{n+1},\ldots,\alpha_{n}+\alpha_{2n}\}$)
and the matrix
$A_{f}=(\alpha_1,\alpha_2,\ldots,\alpha_{2n})_{n\times2n}$ has the
property that any two rows have symplectic inner product zero and
all the rows are independent.
\end{itemize}
\end{lemma}
In the above lemma, the QECC is constructed by giving the projector
$P=f(P_1,P_2,\cdots, P_{n})$ onto the code,  and $P$ is constructed
in the sense of a logic of projection operators given in
Ref\cite{VR07}, where
$\{P_{n+1-i}=\frac{1}{2}(I+E_{\gamma_{i}})|1\leq i\leq n\}$,
$\gamma_{i}$ is the $i-th$ row of the matrix $A_f$. From the
symplectic orthogonality of the rows of $A_{f}$, we have
$\{P_i|1\leq i\leq n\}$ are pairwise commutative, and the
error-correcting ability of the QECC is ensured by the properties of
$Zset_f$. An arbitrary error $e$ acting nontrivially on one qubit
only takes the projector $P_{f_{(x)}}$ to $P_{f(x+t)}$, i.e.,
$eP_{f_{(x)}}e=P_{f(x+t)}$, where $t$ is an element in $Zsetf$, and
$P_{f(x+t)}$ is orthogonal to $P_{f(x)}$.

\section{logic functions and quantum states}

For a logic function $f(x)$ with $n$ variables over $F_p$, it
corresponds to a vector $s=p^{-\frac{n}{2}}\zeta^{f(x)}$, which can
be interpreted as the probability distribution vector of the quantum
state
\begin{equation}
|\psi_{f} \rangle =p^{-\frac{n}{2}}\sum\limits_{x\in
F^{n}_p}\zeta^{f(x)}|x\rangle
 \end{equation}
Specially, if a state has the form of
$2^{-\frac{n}{2}}\sum\limits_{x\in F^{n}_2}(-1)^{f(x)}|x\rangle$,
where $f(x)$ is a Boolean function, we call it a Boolean state
corresponding with $f(x)$.

Then if an error $E_{(a,b)}$ acts on the state$|\psi_{f} \rangle$,
it takes it to another state which is proportional to
$p^{-\frac{n}{2}}\sum\limits_{x\in F^{n}_p}\zeta^{f(x-a)+b\cdot
x}|x\rangle$, which can also be expressed in terms of a logic
function, $f(x)\rightarrow f(x-a)+b \cdot x$.

 Let $K$ Boolean functions be
$g_{i}(x)=f(x)+\beta_{i}\cdot x$, and for $1\leq i<j \leq K$,
$\beta_i\neq \beta_j$ , define $K$ quantum states as
$|\psi_{i}\rangle=p^{-\frac{n}{2}}\sum\limits_{x\in
F^{n}_p}\zeta^{g_{i}(x)}|x\rangle$, then we have the following
theorem.

\begin{theorem}
The subspace spanned by $\{ |\psi_{i}\rangle |1\leq i\leq K \}$ is
an $((n,K,d'))_p$ quantum code, where $d'=min\{ W_s(u,v)|there~
exist~ 1\leq i \leq j \leq K~ such~ that~
W_s(u,v+\beta_{i}+\beta_{j})\geq d \}$, where $d$ is the APC
distance of $f(x)$.
\end{theorem}

\textbf{Proof}.  We only need to prove that for any error
$\varepsilon_{d}$ acting nontrivially on less than $d$ qubits,
$\langle \psi_{i}|\varepsilon_{d}|\psi_{j}
\rangle=f(\varepsilon_{d})\delta_{ij}$ for all $1\leq i,j \leq K$.
Without lose of generosity we assume that $\varepsilon_{d}=\mathcal
{X}_{u}\mathcal {Z}_{v}$ for some pair of vectors in $F^{n}_p$ with
$W_s(u,v)<d'$.

We have
\begin{equation}
\langle \psi_{i}|\varepsilon_{d}|\psi_{j}\rangle  \propto
\sum\limits_{x\in F^{n}_p}\zeta^{f(x-u)+(\beta_{i}+\beta_{j}+v)\cdot
x+f(x)}.
\end{equation}

 Since $W_s(u,v+\beta_{i}+\beta_{j}) < d$, so
$\langle \psi_{i}|\varepsilon_{d}|\psi_{j}\rangle$=0.

Then we verify  that $\langle \psi_{i}|\varepsilon_{d}|\psi_{i}
\rangle$ only depends on $\varepsilon_{d}$.
\begin{equation}
\langle\psi_{i}|\varepsilon_{d}|\psi_{i} \rangle \propto
\sum\limits_{x\in F^{n}_p} \zeta^{f(x-u)+v\cdot x+f(x)}=0
\end{equation}

So, $\{ |\psi_{i}\rangle |1\leq i\leq K \}$ span an $((n,K,d'))_p$
quantum code.$\sharp$

\section{logic functions and graphical QECCs }
\subsection{quadratic Boolean functions  of the form $\frac{1}{2}x\Gamma
x^{T}$}
 Now we consider a class of quadratic logic function
corresponding with a simple undirected graph.

 If $f(x)$ is a quadratic Boolean
function and can be represented as $f(x)=\frac{1}{2}x\Gamma
x^{T}$,where $\Gamma_{n\times n}$ is a symmetric matrix with
elements in $F_{2}$ and vanishing diagonal entries, then the state
$|\psi_f\rangle$ can be viewed as a graph state because $\Gamma$ can
be viewed as the adjacency matrix of a graph $G=(V,E)$, where $V$
and $E$ denote the set of vertices and edges respectively and
$|V|=n$. If we label every vertex of the graph $G$ of $n$ vertexes
from 1 to $n$, then every vertex corresponds with one qubit, and the
error $\varepsilon_{d}=\mathcal {X}_{u}\mathcal {Z}_{v}$ can be
written as $\mathcal {X}_{\omega}\mathcal {Z}_{\delta}$, where
$\omega,\delta$ are subsets of $V$.

Consider the operator $\mathcal {G}_{a}=\mathcal
{X}_{a}\prod\limits_{b\in N_{a}}\mathcal {Z}_{b}$, where $N_{a}$
represents the neighborhood of the vertex $a$ and is denoted by
$N_{a}=\{v\in V|\Gamma_{av}=1\}$, and it was shown in
Ref\cite{MJH04}
 that $\mathcal
{G}_{a}|\psi_f \rangle=|\psi_f \rangle$, so $\mathcal
{X}_{\omega}\mathcal {Z}_{N_\omega}|\psi_f \rangle=|\psi_f \rangle$.

Choose $K$  subsets of $V$ $\{C_{i}|1\leq i \leq K\}$, we define $K$
pair-wise orthogonal quantum states\cite{SQC07} as
\begin{equation}
|\psi_{i}\rangle=2^{-\frac{n}{2}}\sum\limits_{x\in
F^{n}_2}(-1)^{g_{i}(x)}|x\rangle=\mathcal {Z}_{C_{i}}|\psi\rangle.
\end{equation}
where  and corresponds to a vector $\beta_i$ of length $n$,
therefore $g_i(x)$ can be expressed as $g_i(x)=f(x)+\beta_i\cdot x$.
It was shown that \cite{SQC07} $\mathcal {G}_{a}|\psi_{i}
\rangle=-|\psi_{i} \rangle$ if $a\in C_{i}$ and $\mathcal
{G}_{a}|\psi_{i} \rangle=|\psi_{i} \rangle$ if $a\in C_{i}$
otherwise.

\begin{definition}\cite{SQC07}
The $d-$uncoverable set $\mathbb{D}_{d}$ that contains all the
subsets of $V$ which can't be covered by less than $d$ vertices is
denoted by
\begin{center}
$\mathbb{D}_{d}=2^{V}-\{\delta \triangle N_{\omega}||\omega \bigcup
\delta | <d\}$
\end{center}
where $\omega \triangle \delta$ denotes the symmetric difference of
two sets $\omega , \delta$, i.e., $\omega \triangle \delta=\omega
\cup \delta-\omega \cap \delta$, and $N_{\omega}$ denotes the
neighborhood of the $\omega$, i.e., for every element $v$ in
$N_{\omega}$, there exist an element $v'$ in $\omega$ such that
$\Gamma_{vv'}=1$.
\end{definition}

\begin{corollary}
If $C=\{C_1,C_2,\ldots,C_{K} \}$ satisfies the following two
conditions,

\begin{center} (1) $\O \in  C$;
 (2) $C_{i}\triangle C_{j} \in \mathbb{D}_{d}$.
\end{center}
  then the subspace
spanned by the basis$\{|\psi_{i}\rangle=\mathcal
{Z}_{C_{i}}|\psi\rangle|1\leq i \leq K\}$ is an $((n,K,d))$ code,

\end{corollary}
\textbf{Proof}. We choose a Boolean function
$f(x)=\frac{1}{2}x\Gamma x^{T}$ with APC distance $\overline{d}$,
and from Theorem 1, $\{|\psi_{i}\rangle|1\leq i \leq K\}$ span a
$((n,K,d))$, where $d=min\{ W_s(u,v)|there~ exist~ 1\leq i \leq j
\leq K~ such~ that~ W_s(u,v+\beta_{i}+\beta_{j})\geq \overline{d}~
\}$.

For an correctable error $\varepsilon_{d}=\mathcal {X}_{u}\mathcal
{Z}_{v}=\mathcal {X}_{\omega}\mathcal {Z}_{\delta}$, $W_s(u,v)\leq
d-1$, we have

\begin{eqnarray}
\langle\psi_{j}|\varepsilon_{d}|\psi_{i}\rangle &{}&\propto
\langle\psi|\mathcal {Z}_{\delta \triangle N_{\omega}}\mathcal
{Z}_{C_i\triangle C_j}|\psi \rangle \\&& \propto \sum\limits_{x \in
F_{2}^{n} }(-1)^{f(x)+f(x+u)+( v+\beta_{i}+\beta_{j} )\cdot x}
\end{eqnarray}

And $W_s(u,v+\beta_{i}+\beta_{j})< \overline{d}$, so
\begin{center}
$\sum\limits_{x \in F_{2}^{n} }(-1)^{f(x)+f(x+u)+(
v+\beta_{i}+\beta_{j} )\cdot x}=0$.
\end{center}
Then we have $\delta \triangle N_{\omega}\neq  C_i\triangle C_j$, in
other words, $C_i\triangle C_j \in \mathbb{D}_{d}$.$\sharp$

 It should be noted here that in Ref\cite{SQC07}  ,the
authors gave three conditions  for the existence of an $((n,K,d))$
over $F_2$ quantum code, and we consider them unnecessary.

\subsection{quadratic logic functions  of the form $\frac{1}{2}(c,x) A (c,x)^{T}$}

Consider an $(n+k)\times (n+k)$ symmetric matrix $A$ with elements
in $F_{p}$ and vanishing diagonal entries, then for every vector $c$
of length $k$ with elements in $F_{p}$, $f(c,x)=\frac{1}{2}(c,x) A
(c,x)^{T}$ is a logic function of $n$ variables, where
$x=(x_1,x_2,\cdots,x_n)$ is a vector of $n$ variables. Notice that
the degree of $f(x)$ is at most two.

Now, we consider the sufficient conditions for the set of states
$\{|\psi_i\rangle=p^{-\frac{n}{2}}\sum\limits_{x\in F^{n}_p}
\zeta^{f(c_i,x)}|x \rangle | c_i\in F_p^k \}$ that can span an
$[\kern-0.15em[n,k,d]\kern-0.15em]_p$ quantum code, i.e., the
required properties of the Boolean function. Basing on Lemma 1,we
have the following theorem.

\begin{theorem}
Suppose $C,X$ are two disjoint sets. $|C|=k,|X|=n,d\geq 2$,
$A=(a_{ij})_{i,j\in C\cup X}$ is a symmetric matrix with elements in
$F_p$ and vanishing diagonal entries. For arbitrary $E\subseteq
X,|E|=d-1$, $I=X \setminus E $, if the rows of $A_{EI}$ are linear
independent , and
\begin{equation}
A_{IC}d^C+A_{IE}d^E=O^I
\end{equation}
implies that

\begin{equation}
d^C=O^C
\end{equation}
. Then the subspace spanned by $\{|\psi_i \rangle\}$ is an
$[\kern-0.15em[n,k,d]\kern-0.15em]_p$ code over $F_p$, where $\zeta$
is a $p-$th primitive root of 1.
\end{theorem}

\textbf{Proof}. We first prove that if $ A_{IC}d^C+A_{IE}d^E=O^I $
where $I=X \setminus E $ implies that $ d^X=O^X$ , then for
different $i,j$, $\langle\psi_j|\psi_i\rangle=0$.

\begin{equation}
\langle\psi_j|\psi_i\rangle \propto \sum\limits_{x\in F^{n}_p}
\zeta^{f(c_i,x)-f(c_j,x)}
\end{equation}
Since $A$ can be expressed as $A=\left( {\begin{array}{*{20}c}
   A_{CC} & A_{CX}  \\
   A_{XC} & A_{XX}  \\
\end{array}} \right)$, then

\begin{equation}
\sum\limits_{x\in F^{n}_p}\zeta^{f(c_i,x)-f(c_j,x)}\propto
\sum\limits_{x\in F^{n}_p}\zeta^{xA_{XC}(c_i-c_j)^T}
\end{equation}
 so
$\langle\psi_j|\psi_i\rangle\neq 0$ iff $A_{XC}(c_i-c_j)^T=O^X$.

Seeking a contradiction, we suppose $A_{XC}(c_i-c_j)^T=O^X$, then
there exist $I\subseteq X$ such that $A_{IC}(c_i-c_j)^T=O^I$. Let
$d^E=O^E$, then $A_{IC}(c_i-c_j)^T+A_{IE}d^E=O^I$, which satisfies
Eq.(13), so $(c_i-c_j)^T=O^C$ which is impossible because of $i\neq
j$. We come to the result that $\{|\psi_i\rangle\}$ span a subspace
of dimension $p^k$.

Then we prove the subspace spanned by $\{|\psi_i\rangle\}$ is an
$[\kern-0.15em[n,k,d]\kern-0.15em]_p$ quantum code, i.e., for an
error $\varepsilon_{d}=\mathcal {X}_a \mathcal {Z}_b$ with
$W_s(a,b)\leq d-1$,
$\langle\psi_j|\varepsilon_{d}|\psi_i\rangle=f(\varepsilon_{d})\delta_{ij}$.

\begin{equation}
\langle\psi_j|\varepsilon_{d}|\psi_i\rangle\propto \sum\limits_{x\in
F^{n}_p}\zeta^{bx+xA_{XC}(c_i-c_j)^T+aA_{XX}x^T}
\end{equation}
Let $\varepsilon_{d}$ acts on qubits corresponding with a subset
$E\subseteq X$, and $a_E,b_E$ are vectors of length $d-1$. For
simplicity, we denote variables in $I$ as $y$, variables in $E$ as
$z$. Then

\begin{equation}
\langle\psi_j|\varepsilon_{d}|\psi_i\rangle \propto
\sum\limits_{x\in F^{n}_p}\zeta^{b_Ez+zA_{EC}(c_i-c_j)^T
+zA_{EE}a_E^T+yA_{IC}(c_i-c_j)^T+yA_{IE}{a_E}^T}
\end{equation}

For different $i,j$, consider linear terms of $y$, if
$A_{IC}(c_i-c_j)^T+A_{IE}{a_E}^T\neq O^I$, then
$\langle\psi_j|\varepsilon_{d}|\psi_i\rangle=0$. If
$A_{IC}(c_i-c_j)^T+A_{IE}{a_E}^T= O^I$, which satisfies Eq.(13), so
$(c_i-c_j)^T=O^C$ which contradicts the fact that $c_i,c_j$ are
different.

Then we verify $\langle \psi_{i}|\varepsilon_{d}|\psi_{i} \rangle$
only depends on $\varepsilon_{d}$.
\begin{center}
$\langle\psi_{i}|\varepsilon_{d}|\psi_{i} \rangle \propto
\sum\limits_{x\in F^{n}_p}\zeta^{b_Ez+zA_{EE}a_E^T+yA_{IE}{a_E}^T}$
\end{center}

Since the rows of $A_{EI}$ are independent, then if
$A_{IE}{a_E}^T=0$, we can know $a_E=0$, thus
$\langle\psi_{i}|\varepsilon_{d}|\psi_{i} \rangle \propto
\sum\limits_{x\in F^{n}_p}\zeta^{b_Ez}$. Because $b_E\neq 0$,
$\langle\psi_{i}|\varepsilon_{d}|\psi_{i}\rangle=0$.

So, $\{ |\psi_{i}\rangle |1\leq i\leq 2^k \}$ span an
$[\kern-0.15em[n,k,d]\kern-0.15em]_p$ quantum code. $\sharp$


\subsection{Graphical $[\kern-0.15em[n,0,d]\kern-0.15em]$ QECC}
Consider the adjacency matrix($n\times n$) $\Gamma$ of a graph, then
the rows of $A=(\omega I|\Gamma)$ can span an self-dual additive
code $C$ over $GF(4)=\{0, 1, \omega, \omega^2\}$, where
$\omega^2+\omega+1=0$. And $C$ is equivalent to a graph code
$D$\cite{MN}.

Let $\alpha_{i},\beta_i$ are the $i-$th column of $I$ and $\Gamma$
respectively, in fact, $\mathcal
{X}_{\alpha_i}\mathcal{Z}_{\beta_i}$ are stabilizers of $D$, if we
can find a Boolean function $f(x)$ satisfying the following
equations:
\begin{equation}
f(x+\alpha_i)=\beta_ix,\  {\rm for\ } 1\leq i\leq n
\end{equation}
then we can state that
$|\psi\rangle=2^{-\frac{n}{2}}\sum\limits_{x\in
F_2^{n}}(-1)^{f(x)}|x\rangle$ is the basis state. So we find that
the graph code is equivalent to $[\kern-0.15em[n,0,d]\kern-0.15em]$,
where $d$ is the APC distance of $f(x)$.

\begin{example}
Consider a complete graph of 4 vertices, then matrix
\begin{center}
$A=\left( {\begin{array}{*{20}c}
   \omega& 0 & 0 & 0 & 0 & 1 & 1 & 1  \\
   0 & \omega & 0 & 0 & 1 & 0 & 1 & 1 \\
   0 & 0 & \omega  & 0 & 1 & 1 & 0 & 1 \\
   0 & 0 & 0 & \omega & 1 & 1 & 1 & 0 \\
\end{array}} \right)$
\end{center}
After computation, we find that
$f(x)=x_1x_2+(x_1+x_2)(x_3+x_4)+x_3x_4$, the APC distance of $f(x)$
is 2, so $\{ |\psi \rangle=2^{-2}\sum\limits_{x\in
F_2^{4}}(-1)^{f(x)}|x \rangle \}$ span a
$[\kern-0.15em[4,0,2]\kern-0.15em]$ QECC.
\end{example}

In fact, under the correspondence defined in Ref\cite{MN}, every
simple undirected graph with adjacency matrix $\Gamma_{n\times n}$
corresponds to an $[\kern-0.15em[n,0,d]\kern-0.15em]$ QECC over
$F_2$ with the basis state $ |\psi
\rangle=2^{-\frac{n}{2}}\sum\limits_{x\in F_2^{n}}(-1)^{f(x)}|x
\rangle $, where $f(x)=\frac{1}{2}x\Gamma x^T$.


\section{Boolean functions and projectors}
In Ref\cite{VR07}, the authors constructed quantum error correcting
codes via the tools of projectors and Boolean functions. They first
redefine a logic of projectors, then on the assumption that they can
construct a certain matrix, which satisfies some properties
corresponding to the $Z_{setf}$ of a Boolean function $f$, finally
they construct a projector onto a quantum code.

In this section, we refine the projectors another logic of
projectors, and come to the result that in some conditions, the
quantum codes under the construction which is similar to that given
Ref\cite{VR07} has Boolean basis states. Now we give our definition
of operator , which is denoted by $E'_{(a,b)}$.

\begin{definition}
Operators $E'_{(a,b)}$ associated with binary vectors $(a,b)\in
F_2^{2n}$ are redefine by
\begin{equation}
E'_{(a,b)}=e'_1\otimes \ldots \otimes e'_{n}
\end{equation}
where $e'_{j}=i^{a_jb_j}e_{j}$ for $1\leq j\leq n$. In other words,
$E'_{(a,b)}=\mathcal {X}_a \mathcal {Z}_b$.

\end{definition}

Basing on the definition of the logic of projection operators in
Ref\cite{VR07}, we define another logic as the following definition.
\begin{definition}
Let $P=\mathcal {X}_a \mathcal {Z}_b,P'=\mathcal {X}_{a'} \mathcal
{Z}_{b'},P''=\mathcal {X}_{a''} \mathcal {Z}_{b''}$ are three
projection operators, where $a,a',a'',b,b',b''$ are vectors of
length $n$. Then we define $P\vee P'=\mathcal {X}_a \mathcal {Z}_b
+\mathcal {X}_{a'}\mathcal {Z}_{b'}$,
 $P\wedge P'=(-1)^{a'b}\mathcal {X}_{a+a'} \mathcal {Z}_{b+b'}$, $\tilde{P}=I-P$ and
 $(P\vee P')\wedge P''=(-1)^{a''b}\mathcal {X}_{a+a''} \mathcal {Z}_{b+b''}+
 (-1)^{a''b'}\mathcal {X}_{a'+a''} \mathcal {Z}_{b'+b''}$.
\end{definition}

\begin{definition}\cite{VR07}
Given an arbitrary Boolean function $f(x_1,x_2,\cdots,x_n)$, we
define the Projection function $\hat{f}(P_1,P_2,\cdots, P_{n})$ in
which $x_i$ is replaced by $P_i$, multiplication, summation and not
operation in Boolean logic are replaced by the meet, join and
 tilde operation in the projection logic described in Definition 7 respectively.
\end{definition}

We denote $P_i^{c_{i}}$ as $P_i$ if $c_i=0$, and $\tilde{P_i}$ if
$c_i=1$.

If we can construct matrix $A_f=(A|B)$ as in Lemma 3, where $A$ and
$B$ are blocks of $A_f$ of size $n\times n$ with the i-th row
vectors $\alpha_i,\beta_i$, corresponding with a Boolean function
$f(x)$ with $n$ variables, then we redefine the operation operator
$P_{n+1-i}=\frac{1}{2}(I+E'_{\gamma_i})$. The projector
$\hat{f}(P_1,P_2,\cdots, P_{n})$ is still a projector onto an
$((n,M,2))$ QECC, where $M$ is the Hamming weight of $f(x)$.

The projector $P$ onto a QECC $Q$ has the form
$P=\sum\limits|\psi\rangle \langle \psi|$, where $|\psi\rangle$ run
over all the basis states of $Q$. Without lose of generosity, we
assume the vector $(t_1,t_2,\cdots,t_n)$ is an element of the
support of $f$, in fact, every element in the support of $f$
corresponds to a basis state. Then the term corresponds to
$(t_1,t_2,\cdots,t_n)$ in $P=\hat{f}(P_1,P_2,\cdots, P_{n})$ is
$P_1^{t_1},P_2^{t_2},\cdots, P_{n}^{t_n}$, which can be written as
\begin{equation}
2^{-n}\sum\limits_{d\in F_2^{n}}\sum\limits_{x\in
F_2^{n}}(-1)^{\lambda(d,t)}|x+\sum\limits_{i=1}^{n}d_i
\alpha_i\rangle \langle x|
\end{equation}
, where $\lambda(d,t)=(\sum\limits_{i=1}^{n}d_i\beta_i)
x+\sum\limits_{i=1}^{n}t_id_i+\sum\limits_{1\leq j< k \leq
n}d_{j}d_{k}\alpha_j\beta_k$.

Now we consider properties that the Boolean function
$f_{(t_1,t_2,\cdots,t_n)}$ should have if
$P_1^{t_1},P_2^{t_2},\cdots, P_{n}^{t_n}$ can be written as
$|\psi_{(t_1,t_2,\cdots,t_n)}\rangle\langle
\psi_{(t_1,t_2,\cdots,t_n)} |$, where
$|\psi_{(t_1,t_2,\cdots,t_n)}\rangle $ is a Boolean state
corresponding with $f_{(t_1,t_2,\cdots,t_n)}$.

For simplicity, we write $\breve{f}$ in place of
$f_{(t_1,t_2,\cdots,t_n)}$, $|\breve{\psi}\rangle$ in place of
$|\psi _{(t_1,t_2,\cdots,t_n)}\rangle$), and

\begin{equation}
|\breve{\psi}\rangle\langle \breve{\psi} |\propto \sum\limits_{s\in
F_2^{n}}\sum\limits_{x\in
F_2^{n}}(-1)^{\breve{f}(x)+\breve{f}(x+s)}|x+s\rangle\langle x|
\end{equation}

Then we have the following theorem.

\begin{theorem}
$A$ is invertible, $\breve{f}(x)$ is
 is quadratic and $\breve{f}(x)+\breve{f}(x+\alpha_i) =\beta_i x+t_i$.
\end{theorem}
\textbf{Proof}. Since $s$(in Eq.(22)) and $d$(in Eq.(21)) run over
$F_2^{n}$, we require that $\alpha_1,\alpha_2,\cdots,\alpha_n$ are
linear independent, and
\begin{equation}
\breve{f}(x)+\breve{f}(x+\alpha_i) =\beta_i x+t_i.
\end{equation}
Then for arbitrary $d\in F^n_2$,
$\breve{f}(x)+\breve{f}(x+\sum\limits_{i=1}^{n}d_i\alpha_i)
=(\sum\limits_{i=1}^{n}d_i\beta_i)
x+\sum\limits_{i=1}^{n}d_it_i+\sum\limits_{1\leq j< k \leq
n}d_{j}d_{k}\alpha_j\beta_k$, which coincides with $\lambda(d,t)$.

Since the right part of Eq.(23) is an affine Boolean function, we
know that $\breve{f}(x)$
 is quadratic.
$\sharp$

If a QECC has Boolean states, then the study of the QECC can again
be converted to the study of Boolean functions corresponding with
them, we say it is possible.

\begin{example}
Define a Boolean function
$g(y_1,y_2,y_3,y_4)=(y_1+y_2+y_3)(y_1+y_2+y_4)$, then $g(y)$ is
partially bent, and $|Supp(g)|=4$, then for every $a\in F_2^4$ with
$r_g(a)=0$ is in $Zsetg$, and
$Supp(g)=\{s_1=(1000),s_2=(0100),s_3=(0011),s_4=(1111)\}$. Let $a_i$
be a unitary vector of length $4$ with 1 in the $i-$th coordinate
and 0 elsewhere. Since $g(y+a_i)=g(y)+b_ix+c_i$($1\leq i\leq
4,~c=(1100)$), we construct the matrix $A_g$ as
\begin{center}
$A_g=\left( {\begin{array}{*{20}c}
   1 & 0 & 0 & 0 & 0 & 0 & 1 & 1  \\
   0 & 1 & 0 & 0 & 0 & 0 & 1 & 1 \\
   0 & 0 & 1 & 0 & 1 & 1 & 0 & 1 \\
   0 & 0 & 0 & 1 & 1 & 1 & 1 & 0 \\
\end{array}} \right)$
\end{center}

where the $i-$th row of $A_g$ is $v_i=(a_i,b_i)$. We can easily
verify that all the rows of $A_g$ are independent and any two rows
have symplectic product zero because the right four columns of $A_g$
form a symmetric matrix. Express $A_g$ as
$A_g=[x_1,x_2,\cdots,x_{8}]$, then for every $\omega$ with
$W_s(\omega)\leq 1$, $A_g\ast \omega^T$ is in $Zsetg$. After
computation, we have
$\breve{f}_1=g+y_2,\breve{f}_2=g+y_1,\breve{f}_3=g+y_1+y_2+y_4+y_3,\breve{f}_4=g+y_3+y_4$.
We can see that for different $i,j$, $\breve{f}_i-\breve{f}_j$ are
linear terms. And $\{2^{-2}\sum\limits_{x\in
F_2^{4}}(-1)^{\breve{f}_i(x)}|x\rangle\}$ spans a
$[\kern-0.15em[4,2,2]\kern-0.15em]$ code, which meets the quantum
singleton bound, and therefore is an MDS code.

\end{example}

In fact, for every function $f(y)$ with $2m$ variables of the form
$f(y)=(y_1+y_2+\cdots+y_{2m-2}+y_{2m-1})(y_1+y_2+\cdots+y_{2m-2}+y_{2m})$
(which is a partially bent function\cite{C93}), we can find
$|Supp(f)|$ Boolean functions $\breve{f}_i$ satisfying that
$\{2^{-m}\sum\limits_{x\in
F_2^{2m}}(-1)^{\breve{f}_i(x)}|x\rangle\}$ spans a
$[\kern-0.15em[2m,2m-2,2]\kern-0.15em]$  MDS code.

Because a quantum code with Boolean basis state is interesting, it
is natural to question what kind of properties of the Boolean
functions
 used in Lemma 3 should satisfy.

\begin{lemma}\cite{CD,FN}
A Boolean function $f(x)$ with $n$ variables is bent if and only if
$r_f(s)=\left\{ \begin{array}{l}
 2^n,\ \ \ s=0 \\
 0,\ \ \ else
 \end{array} \right.$, and if $f(x)$ is a bent function, then
 $|Supp(f)|=2^{n-1}\pm 2^{n/2-1}$.
\end{lemma}

\begin{theorem}
For arbitrary $(c_1,c_2,\cdots,c_n)\in Supp(f)$ , the Boolean
function $f$ with more than 2 variables used in Lemma 3 can't be a
bent function.
\end{theorem}
\textbf{Proof}. Seeking a contradiction, we assume that $f$ is bent,
then for every $s\in F_2^n,~s\neq 0$, $r_f(s)=0$. If $Zsetf\neq
\varnothing$, then for every $a\in Zsetf$, $r_f(a)=0$. From Lemma 2,
the weight of $f$ is equal to $2^{n-2}$, which contradicts the
property of bent functions described in Lemma 4, so $f$ is not bent.
$\sharp$

\section*{Acknowledgment}

This work is supported by the Natural Science Foundation of China
under Grant No. 60403004, the Outstanding Youth Foundation of Henan
Province under Grant No.0612000500.

The authors would like to thank Markus Grassl for helpful and
important discussion.

\end{document}